\documentclass[prl,superscriptaddress,showpacs,amssymb,amsmath,
amsfonts,aps,twocolumn,secnumarabic,floatfix]{revtex4}
\usepackage{graphics}
\usepackage{epsfig}
\begin{document} 
\bibliographystyle{try} 

\topmargin -0.9cm 
 
 \title{Unitarity constraints on neutral pion electroproduction}

\newcommand*{\JLAB }{ Thomas Jefferson National Accelerator Facility, Newport News, Virginia 23606} 
\affiliation{\JLAB } 

\author{J.M.~Laget}
     \affiliation{\JLAB}

\date{\today} 

\begin{abstract} 

At large virtuality $Q^2$, the coupling to the vector meson production channels provides us with a natural explanation of the surprisingly large cross section of the neutral pion electroproduction recently measured at Jefferson Laboratory, without destroying the good agreement between the Regge pole model and the data at the real photon point. Elastic rescattering of the $\pi^0$ provides us with a way to explain why the node, that appears  at $t\sim -0.5$~GeV$^2$ at the real photon point, disappears as soon as $Q^2$ differs from zero. 
%It seems that we have not yet reached the domains of factorization. 
\end{abstract} 
 
 
\maketitle 

The electroproduction of neutral pion raises two issues. The first one is a long standing issue (mid 70's): the node that appears at $t\sim$ -0.5~GeV$^2$ at the real photon point~\cite{An70} disappears as soon as the virtuality of the photon $Q^2$ differs from zero~\cite{Bra78}. Several attempts, ranging from Regge cuts~\cite{Ha71,Gol73,Ahm08} to direct coupling to quarks and rearrangement~\cite{Nat76}, were proposed to qualitatively explain this result. However, no quantitative explanation has been proposed yet, except at the expense of a strong variation with $Q^2$ of the elatic cut~\cite{Col81}. The second issue is more recent and was raised, about three years ago, during the preliminary analysis of the data obtained at Jefferson Laboratory (JLab) in HallA and in HallB. When extrapolated at large $Q^2$, any Regge pole model that leads to a fair understanding  of the low $Q^2$ data grossly underestimates the JLab data. The HallA data~\cite{Cam10} have just been released; the HallB data~\cite{Kub10} are still in the final stage of analysis and will be released soon.

In this Letter, I propose a solution of these two problems which combines elastic $\pi^0$ rescattering and the coupling to inelastic $\omega p$, $\rho^+ n$, $\rho^+ \Delta^0$ and $\rho^- \Delta^{++}$ channels. A recent measurement at JLab~\cite{Fra09} has shown that the cross section of the  $p(\gamma^*,\rho^+) n$ channel, which is very small at the real photon point, becomes comparable to, and even larger than, the cross section of the $p(\gamma^*,\rho^0) p$ channel, at large virtuality $Q^2$.

\begin{figure}[hbt]
\begin{center}
\epsfig{file=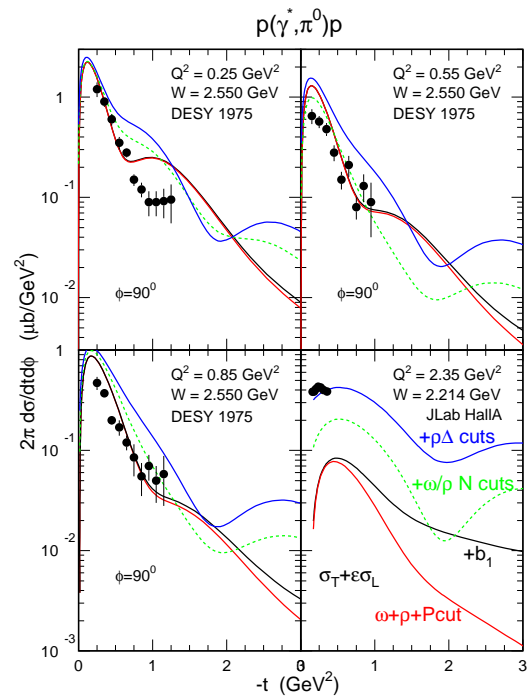,width=3.in}
\caption[]{(Color on line) The cross section of the $p(\gamma^*,\pi^0)p$ reaction recorded at DESY~\cite{Bra78}, top and bottom left, and JLab~\cite{Cam10}, bottom right. The basic Regge Pole model with the Pomeron cut corresponds to the red (without $b_1$ pole) and the black (with $b_1$ pole)  full line curves. The dashed line curves (green) take into account the contribution of the $\pi$  charge exchange scattering cuts, the inelastic $\omega p$ and $\rho^+ n$ cuts. The (blue) full line curves take also into account the contribution of the inelastic $\rho^{\pm} \Delta$ cuts. }
\label{dsdt_pizero}
\end{center}
\end{figure}

Figure~\ref{dsdt_pizero} summarizes these findings. The DESY data have been recorded in a geometry where the $\pi^0$ is emitted perpendicular to the electron scattering plane (azimuthal angle $\phi = 90^{\circ}$) that emphasizes the transverse component of the cross section. The JLab data have been integrated over the $\pi^0$ azimuthal angle and correspond to the combination of the transverse and longitudinal cross sections $d\sigma_T/dt + \epsilon d\sigma_L/dt$, where $\epsilon$ is the polarization of the virtual photon. The combination of the Regge pole and the $\pi^0$ elastic scattering amplitude (Pomeron cut) allows to reproduce the cross section at moderate $Q^2$ ($\sim 0.5 \div 1$~GeV$^2$), but underestimates the JLab data at higher $Q^2$. The coupling to the charged $\rho$ production channels brings the theoretical cross section close to the JLab data at $Q^2=$ 2.3~GeV$^2$. It is remarkable, and was rather unexpected, that the cross section, at $-t\sim$ 0.3, does not change between $Q^2=$ 0.85 and 2.3 GeV$^2$.

The basic Regge pole amplitudes are fully described in ref.~\cite{Gui97b} (GLV). Instead of their expression in terms of $\gamma$ matrices, I use the expression of the  $\omega$ and $\rho$ (that are given in the appendix of ref.~\cite{La06}), as well as the expression of the $b_1$, $t$-channel exchange amplitudes in terms of $\sigma$ matrices. In the GLV scheme, the node in the real photon cross section near $t=$ -0.5~GeV$^2$ is naturally generated by the use of a non-degenerated amplitude of the $\omega$ Regge pole. The contribution of the $\rho$ Regge pole with a degenerated amplitude fills in the dip and brings the GLV model close to the data. The exchange of the $b_1$ meson does not contribute significantly to the unpolarized cross section, but is needed to account for the  measured photon asymmetry $\Sigma$. Multiplying each amplitude by an electromagnetic form factor $F_M(Q^2)$ keeps the shape of the real photon cross section in the virtual photon sector, in contradiction with experiment.

The other way to generate a node in a Regge amplitude  is to use a degenerated amplitude  and supplement it  by the elastic absorptive cut. Assuming that the elastic scattering amplitude is driven by the Pomeron exchange, it is possible to express the cut amplitude as an effective Regge pole~\cite{Do02}. In this scheme, the $\omega$ Regge exchange amplitude takes the form:
\begin{eqnarray}
{\cal T}_{\omega} &=& {\cal T}_F \times (t-m_{\omega}^2)\times
\frac{g_{\omega NN}}{g_{\omega NN}^{GLV}}
\nonumber \\
&&\left( e^{-i\pi \alpha_{\omega}(t)}
\left(  \frac{s}{s_0} \right)^{\alpha_{\omega}(t)-1} 
\alpha'_{\omega} \Gamma(1-\alpha_{\omega})F_{\omega}(Q^2)
\right. \nonumber \\ && \left. 
- e^{-i\pi \delta_{c}(Q^2)}
\left(  \frac{s}{s_0} \right)^{\alpha_{c}(t)-1} 
\alpha'_{c} \Gamma(1-\alpha_{c})F_{c}(Q^2) G_c(t)
\right)  
\nonumber \\
\label{T_Regge}
\end{eqnarray}
where $t=(k_{\omega}-k_{\gamma})^2$ is the four momentum transfer and $m_{\omega}$ is the mass of the $\omega$ meson. The Feynman amplitude ${\cal T}_F$ has the same spin-momentum structure as in the GLV scheme. Only the $g_{\omega NN}$ coupling constant will be re-fitted to experiment. The $\omega$ Regge trajectory is the same as in the GLV scheme:
\begin{eqnarray}
\alpha_{\omega}(t)&=& \alpha_{\omega}(0)+ \alpha'_{\omega} t
= 0.44 +0.9t
\label{omega_traj}
\end{eqnarray}
while the intercept and the slope of the effective trajectory of the cut take the form:
\begin{eqnarray}
\alpha_{c}(0)&=& \alpha_{\omega}(0)+ \alpha_{P}(0)-1\;\;\;\;\;\, = 0.44
\nonumber \\
\alpha'_c
&=& (\alpha'_{\omega}\times\alpha'_{P})/(\alpha'_{\omega}+\alpha'_{P})
= 0.2
\label{cut_traj}
\end{eqnarray}
where the intercept and the slope of the Pomeron Regge trajectory are respectively $\alpha_P(0)=1$ and $\alpha'_P= 0.25$

\begin{figure}[hbt]
\begin{center}
\epsfig{file=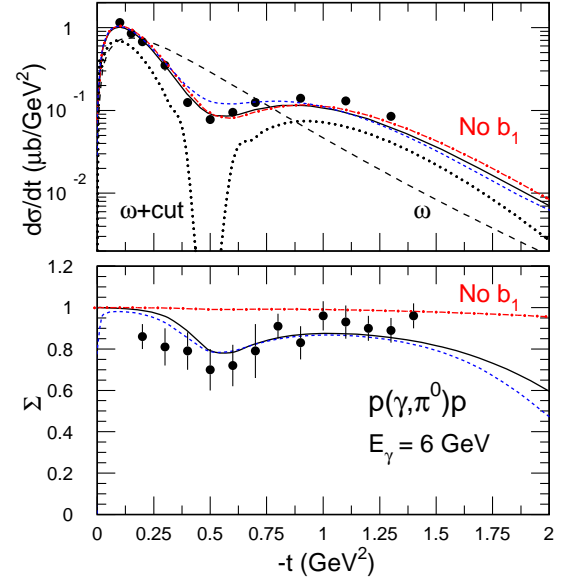,width=3.in}
\caption[]{(Color on line) The cross section of the $p(\gamma,\pi^0)p$ reaction~\cite{An70} (top) and the photon asymmetry~\cite{An71} (bottom). The dashed line curve corresponds to the  $\omega$ Regge Pole amplitude with a degenerated trajectory. The dotted line curve includes the Pomeron cut. The red dot-dashed line curves include the $\rho$ exchange, while the black full lines curves include also the $b_1$ exchange.  The blue short-dashed line curves include all the inelastic cuts.}
\label{P_cut_6GeV}
\end{center}
\end{figure}

The purely destructive interference between the pole and the cut amplitudes is a direct consequence of the structure of the rescattering loop amplitude and the almost purely absorptive nature of the elastic $\pi N$ scattering amplitude (see eqs. (9) to (11) and the corresponding discussion in ref.~\cite{La10}). It is possible to reproduce the GLV non-degenerated $\omega$ amplitude (squared) with the following choice (dotted line curve in Fig.~\ref{P_cut_6GeV}):
\begin{eqnarray}
 g_{\omega NN}^2/4\pi&=& 4.9
\nonumber \\
G_c (t)&=& 3.7 e^{2t}
\label{P_cut}
\end{eqnarray}
It is worth pointing out that the $\omega NN$ coupling constant is about half of the GLV one ($g_{\omega NN}^2/4\pi=$ 17.9), in better agreement with the range of values that are determined in the analysis of low energy $NN$ scattering data. It is also almost the same as the value ($g_{\omega NN}^2/4\pi=$ 6.44)  needed in the analysis of the $p(\gamma,\eta)p$ reaction~\cite{La05}. In this channel, there is no node and the $\eta N$ scattering cross section is much lower than the $\pi N$ one: The use of a degenerated $\omega$ Regge amplitude alone, with no elastic cut, is more justified. The exchange of the $\rho$ degenerated trajectory, with the same coupling constants as in GLV, brings the model close to the unpolarized cross section.

In the first version of this paper~\cite{La10b} an axial-vector $b_1NN$ coupling was used, as in the GLV scheme. It turns out that this coupling is forbiden: it is allowed at the $a_1NN$ vertex, but the $a_1$ meson does not couple to the neutral pion. Therefore, I use the following $b_1$ exchange current which is based on an axial-tensor  $b_1NN$ coupling:
\begin{eqnarray}
{\cal J}^{\mu}_{b_1}&=& \frac{g_{b_1\pi\gamma}}{m_{\pi}} g_{b_1} 
\frac{\kappa_{b_1}}{2m} {\overline u}(p_f) \gamma_5 u(p_i)
(2p^{\mu}_f k p_i -2p^{\mu}_i k p_f)
\nonumber \\ &&
e^{-i\pi \alpha_{\pi}(t)}
 \left( \frac{s}{s_0} \right)^{\alpha_{\pi}(t)-1} 
\alpha'_{\pi} \Gamma(1-\alpha_{\pi})F_{b_1}(Q^2) \;\;
\label{b1_pole}
\end{eqnarray}
where the Regge trajectory is the same as the pion degenerated trajectory and the electromagnetic coupling constant $g_{b_1\pi\gamma}$ is the same as in GLV. The choice of the strong coupling constants of the $b_1$ meson $g^2_{b_1}/4\pi=$~21.5 and $\kappa_{b_1}=$~2 leads to a fair agreement with the beam asymmetry at the real photon point. 

This scheme offers us with a way of shifting the minimum of the cross section when the virtuality $Q^2$ of the photon increases, by using slightly different cut off masses in the electromagnetic form factors of the poles and the Pomeron cut, $F(Q^2)=1/(1+Q^2/{\Lambda ^2})$. The following choice leads to a good accounting of the DESY data at $Q^2=$ 0.85~GeV$^2$: $\Lambda_{\omega}^2=$ 0.325~GeV$^2$, $\Lambda_{\rho}^2=$ 0.400~GeV$^2$, $\Lambda_{b_1}^2=$ 1.~GeV$^2$,  $\Lambda_{c}^2=$ 0.300~GeV$^2$ and $\delta_c(Q^2)= -0.46 Q^2/0.85$.

The agreement is good too at $Q^2=$ 0.55~GeV$^2$, but it is not possible to get rid of the second maximum in $t$ when $Q^2=$ 0.22~GeV$^2$. Also, the extrapolation of this scheme at $Q^2=$ 2.3~GeV$^2$ misses the recent JLab data.

\begin{figure}[hbt]
\begin{center}
\epsfig{file=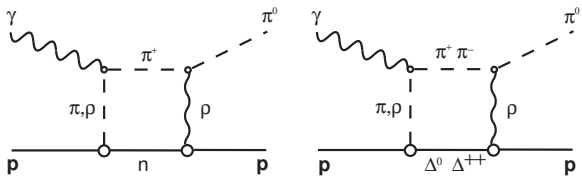,width=3.in}
\caption[]{ The Charge Exchange pion rescattering graphs. }
\label{CEX_graph}
\end{center}
\end{figure}

The first inelastic cut that may play a role is the Charge Exchange (CEX) pion rescattering cut (Fig.~\ref{CEX_graph}). The charged pion electroproduction~\cite{Hub08}, around $Q^2=$ 2.3~GeV$^2$,  is larger ($\sim 3$ $\mu$b/GeV$^2$) than the $\pi^0$ one ($\sim 0.4$ $\mu$b/GeV$^2$) at low $t$. Neglecting its principal part, the corresponding rescattering matrix element reduces to~\cite{La10}:
\begin{eqnarray}
 T_{\pi N}&=& -i\frac{p'_{c.m.}}{16\pi^2}\frac{m}{\sqrt{s}}
 \int  d{\Omega} \left[T_{\gamma^* p\rightarrow \pi^+ n}(t_{\gamma})
 T_{\pi^{+} n \rightarrow \pi^{0} p}(t_{\pi})
 \right]
 \nonumber \\ && 
 \label{pi_cex_cut}
\end{eqnarray}
where $p'_{c.m.}=\sqrt{(s-(m_{\pi}-m)^2)(s-(m_{\pi}+m)^2)/4s}$  is the on-shell momentum  of the intermediate neutron, for  the c.m. energy $\sqrt s$. The two fold integral runs over the solid angle $\Omega$ of the intermediate neutron, and is performed numerically. The four momentum transfer between the incoming photon and the intermediate $\pi$ is $t_{\gamma}=(k_{\gamma}-P_{\pi})^2$, while the four momentum transfer between the intermediate and the outgoing pions is $t_{\pi}=(k_{\pi}-P_{\pi})^2$. The summation over all the spin indices of the intermediate particles is meant.

\begin{figure}[hbt]
\begin{center}
\epsfig{file=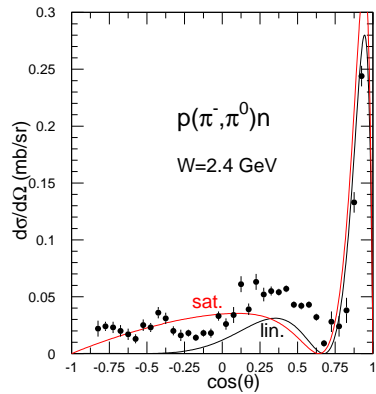,width=2.in}
\caption[]{(Color on line) The $\pi$N CEX at $\sqrt s =$ 2.4~GeV~\cite{Su87}. The (black) line, marked "lin", corresponds to the use of a linear Regge trajectory, while the (red) line, marked "sat", corresponds to the use of a saturating Regge trajectory. }
\label{piN_cex}
\end{center}
\end{figure}

For the $p(\gamma^*, \pi^+)n$ amplitude, I use the VGL model~\cite{Va98} which reproduces fairly well the experimental data~\cite{Hub08} around $Q^2=$2.3~GeV$^2$ and $\sqrt s=$ 2.2~GeV, at least the Longitudinal part. The expression of the CEX amplitude is:
\begin{eqnarray}
{\cal T}_{CEX} &=& \sqrt{\frac{3}{2}}
\frac{g_{\rho}(1+\kappa_V)}{m}g_{\rho\pi\pi} 
{\cal P}_R^{\rho} F_1(t_{\pi})
\nonumber \\
&& \left ( \lambda_f \left | 
\vec{\sigma} \cdot \vec{P_{\pi}}\times \vec{k_{\pi}}
\right |\lambda_i \right )
\label{pi_cex}
\end{eqnarray}
where $g^2_{\rho\pi\pi}/4\pi=$ 5.71, where $g^2_{\rho}/4\pi =$ 0.92, where $\kappa_V=$ 6 and where $F_1$ is the nucleon form factor as defined in~\cite{La10}. Since the experimental $t$ distribution exhibits a node, I use the non degenerated Regge propagator ${\cal P}_R^{\rho}$ with the saturating trajectory of the $\rho$~\cite{Gui97b}. As shown in Fig.~\ref{piN_cex}, this gives a good account of the $\pi$N CEX scattering data~\cite{Su87} in the same energy range.

Since the $\pi N \Delta$ and $\rho N \Delta$ coupling constants are comparable to and even larger than the $\pi NN$ and the $\rho NN$ ones (see for instance ref.~\cite{La81}), the $\pi^+ \Delta^0$ and $\pi^- \Delta^{++}$intermediate states play also a role (Fig.~\ref{CEX_graph}). The rescattering matrix element is a straightforward extension of eqs.~(\ref{pi_cex_cut}) and~(\ref{pi_cex}), using the relevant coupling constants as well as the relevant isospin coefficients, and replacing the $\vec{\sigma}$ matrices by the $N\rightarrow \Delta$ spin transition matrices $\vec{S}$.

\begin{figure}[hbt]
\begin{center}
\epsfig{file=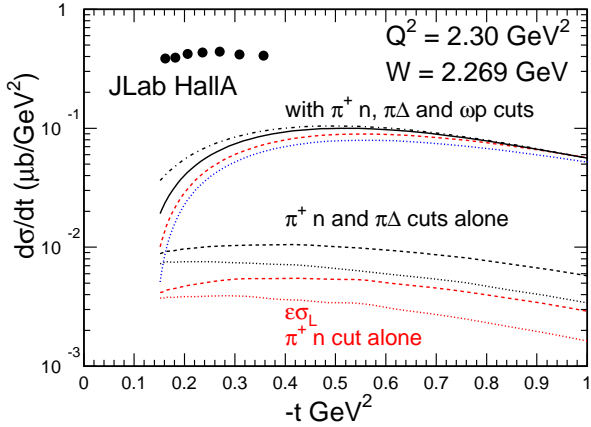,width=3.4in}
\caption[]{(Color on line) The contribution of the $\pi N$ (dashed  red line), the $\pi\Delta$ (full black line) CEX cuts and of the $\omega p$ cut (dash-dotted black line) to the cross section at $Q^2=$ 2.3~GeV$^2$ and $\sqrt s=$ 2.269~GeV. The dotted blue line is the contribution of the Regge poles. The contribution of the $\pi N$ and the $\pi N$ + $\pi\Delta$ cuts alone is shown in the bottom of the figure. The dotted lines correspond to the longitudinal component only.}
\label{dsdt_cex}
\end{center}
\end{figure}

Fig.~\ref{dsdt_cex} shows that the coupling to these CEX channels is not enough to account for the large experimental cross section at $Q^2=$ 2.3~GeV.

\begin{figure}[hbt]
\begin{center}
\epsfig{file=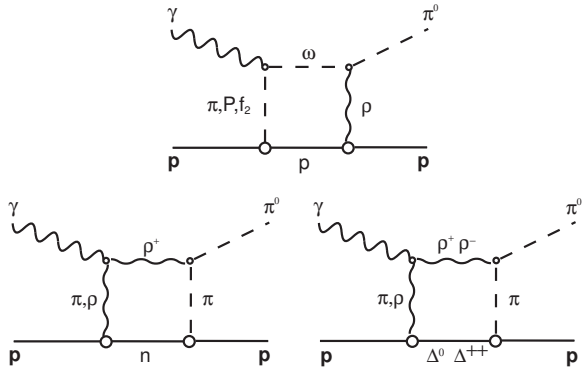,width=3.in}
\caption[]{ The Vector meson cuts. }
\label{vec_graphs}
\end{center}
\end{figure}

The next cuts are the vector meson cuts shown in Fig.~\ref{vec_graphs}. The generic amplitude is:
\begin{eqnarray}
 T_{V N}&=& -i\frac{p_{c.m.}}{16\pi^2}\frac{M}{\sqrt{s}}
 \int  d{\Omega} \left[T_{\gamma p\rightarrow V N}(t_{\gamma})
 T_{V N \rightarrow \pi^{0} p}(t_{\pi})
 \right]
\nonumber \\
 \label{sing}
\end{eqnarray}
where $p_{c.m.}=\sqrt{(s-(m_{V}-M)^2)(s-(m_{V}+M)^2)/4s}$  is the on-shell momentum  of the intermediate baryon (of mass $M$), for  the c.m. energy $\sqrt s$. The two fold integral runs over the solid angle $\Omega$ of the intermediate baryon. The four momentum transfer between the incoming photon and the vector meson is $t_{\gamma}=(k_{\gamma}-P_{V})^2$, while the four momentum transfer between the vector meson and the outgoing pion is $t_{\pi}=(k_{\pi}-P_{V})^2$. The summation over all the spin indices of the intermediate particles is meant. 

\begin{figure}[hbt]
\begin{center}
\epsfig{file=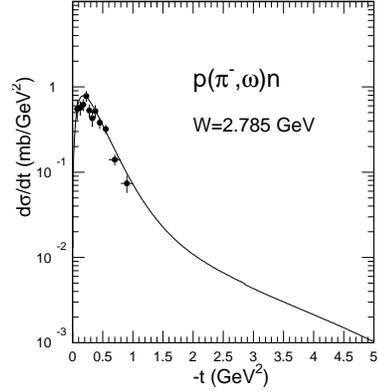,width=2.in}
\caption[]{ The cross section of the  $p(\pi^-,\omega)n$ reaction at $\sqrt s =$ 2.785~GeV~\cite{Ho73}. }
\label{ometopi_cross}
\end{center}
\end{figure}

Since the $\rho^0$ cannot decay into two $\pi^0$'s, only the $\omega p$, the $\rho^+ n$ and the $\rho^{\pm} \Delta$ cuts have to be taken into account. In the $\omega p$ cut, the amplitude of the $p(\gamma^*,\omega)p$ reaction is based on the exchange of the Regge trajectories of the Pomeron, the  $\pi$, the $f_2$ in the $t$-channel and of the proton  in the $u$-channel. The model is described in ref.~\cite{La04} and reproduces well the experimental data~\cite{Mo05} in the JLab energy and momentum range. The amplitude of $p(\omega,\pi^0)p$ has the same structure~\cite{La06} as the $\rho$ exchange part of the Regge amplitude of the reaction $p(\gamma,\pi^0)p$, to which it is related under the Vector Meson Dominance assumption:
\begin{eqnarray}
{\cal T_{\omega\pi}}&=& \frac{f_{\omega}}{\sqrt{4\pi}} \times
{\cal T_{\gamma\pi}}
\label{ometopi0}
\end{eqnarray}
where  $f_{\omega}^2/4\pi=$ 18.4, and where the amplitude is evaluated with the actual kinematics of the $p(\omega,\pi^0)p$ reaction. This model leads to a very good agreement with the available data (Fig.~\ref{ometopi_cross}). Again, the contribution of the $\omega p$ cut is not enough to reproduce the experimental data at $Q^2=$ 2.3 GeV$^2$ (Fig.~\ref{dsdt_cex}).

\begin{figure}[hbt]
\begin{center}
\epsfig{file=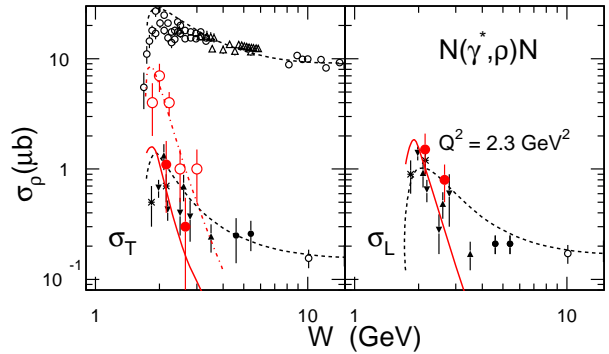,width=3.5in}
\caption[]{ (Color on line) The comparison between the cross sections of the $p(\gamma,\rho^0)p$ (black) and $p(\gamma,\rho^+)n$ (red) reactions at $Q^2=$ 0 (top) and $Q^2=$ 2.3~GeV$^2$ (bottom). The open red circles and the dash-dotted red curve correspond to the photo production of $\rho^-$, while the filled red circles and the full red curves correspond to the electroproduction of $\rho^+$.}
\label{rho_Xsec}
\end{center}
\end{figure}

\begin{figure}[hbt]
\begin{center}
\epsfig{file=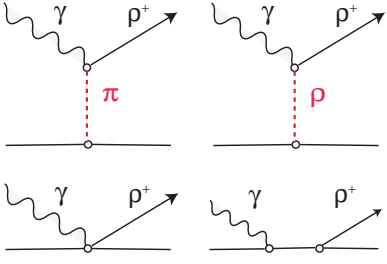,width=2.in}
\caption[]{ The graphs of the reaction $p(\gamma,\rho^+)n$.Top: meson exchange graphs. Bottom: Contact and nucleon exchange gauge graphs. }
\label{graph_rho+}
\end{center}
\end{figure}

The contribution of the $\rho^+n$ cut is far more important (Fig.~\ref{dsdt_pizero}). The first reason is that the cross section of the $N(\rho,\pi)N$ reaction is larger than the cross section of the $N(\omega,\pi)N$ reaction at low $t$ (compare Fig.~3 of~\cite{La10} and Fig.~\ref{ometopi_cross}): The former reaction is driven by $\pi$ exchange while the latter is driven by $\rho$ exchange. The second reason is that the cross section of the $p(\gamma,\rho^+)n$, which is very small at the real photon point, becomes large at large $Q^2$. This is  shown in Fig.~\ref{rho_Xsec} which compares the cross sections of the $p(\gamma,\rho^0)p$ and $p(\gamma,\rho^+)n$ reactions at the real photon point and at $Q^2=$ 2.3~GeV$^2$. The reference to the experimental cross sections of the $\rho^0$, as well as the description of the model,  can be found in~\cite{La04, La00}. At the real photon point, the cross section of the $n(\gamma,\rho^-)p$ reaction comes from~\cite{Be74}. At $Q^2=$ 2.3~GeV$^2$, the two data of the $p(\gamma,\rho^+)n$ reaction are preliminary and given for illustration only. They come from the  preliminary analysis of a JLab experiment~\cite{Fra09}. The model of the $p(\gamma,\rho^+)n$ reaction is based on the exchange of the $\pi^+$ and $\rho^+$ Regge trajectories. The $\pi$ exchange amplitude is the same as the amplitude~\cite{La04,La00} that dominates the $p(\gamma^*,\omega)p$ reaction (with trivial changes of the coupling constants). It is not important and the dominant contribution comes from the $\rho$ exchange amplitude. Its vector part takes the form:
\begin{eqnarray}
{\cal T}_{\gamma\rho^+}&=& \frac{eg_{\rho}(\kappa_V+1)}{2m}\sqrt{2}\; 
F_{\rho}(Q^2,t)\; (t-m_{\rho}^2)\; {\cal P}^{\rho}_{R}
\nonumber \\ && \times
\left(\lambda_f \left |
(2\vec{P_{\rho}}-\vec{k_{\gamma}}) \cdot \vec{\epsilon} \;
\frac{\vec{\sigma}\cdot (\vec{P_{\rho}}- \vec{k_{\gamma}}) 
\times \vec{\epsilon_{\rho}}} {t-m_{\rho}^2}
\right. \right. \nonumber \\ &&
\left. \left. 
+(2\vec{p_i}+ \vec{k_{\gamma}}) \cdot \vec{\epsilon} \;
\frac{\sigma \cdot \vec{P_{\rho}}\times \vec{\epsilon_{\rho}}}{s-m^2}
+ \vec{\sigma} \cdot \vec{\epsilon} \times \vec{\epsilon_{\rho}}
\right | \lambda_i \right )
\nonumber \\
&& \times \frac{\sqrt{(E_i+m)(E_f+m)}}{2m}
\label{rho+_amp}
\end{eqnarray}
where $\vec{\epsilon}$ and $\vec{\epsilon_{\rho}}$ are respectively the polarization vectors of the incoming photon and the outgoing $\rho$, and where the coupling constants are $g^2_{\rho}/4{\pi}=$ 0.4 and $\kappa_V=$ 6. One recognizes easily the $\rho$ pole term as well as the nucleon s-channel pole term and the contact term (see Fig.~\ref{graph_rho+})  that are necessary to make the amplitude gauge invariant.

I use the GLV degenerated Regge propagator ${\cal P}^{\rho}_{R}$, with a linear trajectory, and the $t$ dependent electromagnetic form factor $F_{\rho}(Q^2,t)$ of~\cite{La04} with a cut-off mass of $\Lambda^2_{\rho}=$  0.9~GeV$^2$. Under those assumptions, the model not only predicts the right transverse and longitudinal integrated cross sections (Fig.~\ref{rho_Xsec}) from the real photon point to large $Q^2$, but also their $t$ distribution in the whole range ($1.5<Q^2<4$ GeV$^2$, $2<\sqrt{s}<2.8$ GeV) that has been covered in the JLab HallB experiment~\cite{Fra09}. The model will be compared to the final JLab data in the experimental paper. 

As shown in Fig.~\ref{dsdt_pizero}, the contribution of the $\rho^+ n$ cut accounts for about half of the cross section at $Q^2=$ 2.3~GeV$^2$, modifies little the cross section at $Q^2=$ 0.55 and 0.85~GeV$^2$, but interferes with the Regge pole contribution at $Q^2=$ 0.25~GeV$^2$. At this point of the discussion it is worth to emphasize that no freedom is left in the amplitudes of the cuts, in that sense that the unitarity integral relates on-mass shell elementary amplitudes of the production and absorption of the intermediate mesons. As soon as those amplitudes reproduce the elementary reaction cross sections, the amplitude of each cut is frozen and their relative contribution is driven by the actual relative size of the cross section of the elementary channels. So, this singular part of the rescattering integrals is on solid grounds and consistently relates several channels that have been studied recently at JLab or elsewhere: it simply cannot be overlooked.

The contribution of the principal part of the rescattering integrals is not so well constrained as it requires the knowledge of either the off-mass shell amplitudes, if is calculated by brute force, or their asymptotic behavior, if it is determined by a dispersion relation. This is an open issue, but I simply note that the measured ratio between the teal part and the imaginary part of the amplitude does not exceed 10\% in Compton scattering or  meson-Nucleon elastic scattering at forward angles. So there is good reason to expect that the principal part of the rescattering integrals contributes little.

The third reason why the coupling to the charged $\rho$ production channels is strong is that the contribution of the $\Delta$'s intermediate states (Fig.~\ref{vec_graphs}) is as important as the contribution of the neutron intermediate state. Since there are no measured cross sections of the $N(\gamma^*,\rho^{\pm})\Delta$ reaction in the virtual photon sector, I note that, at lowest order, the structure of the lower part of the loop diagrams is very similar:
\begin{eqnarray}
\vec{\sigma} \cdot \vec{k_{\rho}} \vec{\sigma} \cdot \vec{k_{\pi}}
&=& \vec{k_{\rho}}\cdot \vec{k_{\pi}}
+i \vec{\sigma} \cdot \vec{k_{\rho}} \times \vec{k_{\pi}}
\nonumber \\
\vec{S^{\dagger}} \cdot \vec{k_{\rho}} \vec{S} \cdot \vec{k_{\pi}}
&=& \frac{2}{3}\vec{k_{\rho}}\cdot \vec{k_{\pi}}
+i \frac{1}{3}\vec{\sigma} \cdot \vec{k_{\rho}} \times \vec{k_{\pi}}
\label{spin_amp}
\end{eqnarray}
Where $\vec{S}$ is the spin operator of the $N\rightarrow \Delta$ transition. Assuming that the scalar and vector parts contribute equally:
 \begin{eqnarray}
{\cal T}_n + {\cal T}_{\Delta^0} +{\cal T}_{\Delta^{++}} 
&=&{\cal T}_n \left(1+ \frac{1}{2} \frac{p_{\Delta}m_{\Delta}}{p \;m}
\frac{G_{\rho}G_{\pi}}{g_{\rho}g_{\pi}} 
\left( \frac{1}{6} +\frac{1}{2}   \right)
\right) \nonumber \\
&=& {\cal T}_n \left(1+1.5\frac{p_{\Delta}m_{\Delta}}{p \;m}\right)
\label{Rho_Delta_cut}
\end{eqnarray}
where $p_{\Delta}$ and $p$ are the momenta of the $\Delta$ and the neutron in the $\rho \Delta$ and $\rho^+ n$ loop respectively, and where the ratio of the coupling constant is $G_{\rho}G_{\pi}/g_{\rho}g_{\pi}=$ 4.49 according to~\cite{La81}. The last bracket contains the ratio of the isospin coefficients ($1/2$ in the $\Delta^{++}$ channel and $1/6$ in the $\Delta^0$ channel). For the $\pi \Delta$ cuts, such an estimate is consistent with the numerical evaluation of the integral that is shown in the bottom part of Fig.~\ref{dsdt_cex}. I also note that, at the real photon point, the same kind of estimate predicts a cross section of the $p(\gamma,\rho^+)\Delta^0$ reaction comparable to the cross section of the  $p(\gamma,\rho^+)n$ reaction, in accord with the only experiment~\cite{Ba79} available so far. 

Under those assumptions, the model becomes close to the JLab HallA data at $Q^2=$ 2.3~GeV$^2$  (Fig.~\ref{dsdt_pizero}) without destoying the good agreement at the real photon point (Fig.~\ref{P_cut_6GeV}). It slightly overestimates the DESY data at $Q^2=$ 0.55 and 0.85~GeV$^2$. This can be fixed by a fine tuning of the cut off masses in the electromagnetic form factors of the Regge poles, which I had chosen to reproduce the DESY data without the cut contribution. This will not affect the agreement at the lowest and the highest $Q^2$. However the model already gives a good account of the preliminary JLab HallB data~\cite{Kub10} in the range $1.5<Q^2<4$ Gev$^2$, $2.1<\sqrt{s}<2.8$ GeV, and I postpone such an adjustment until the final data are released.

\begin{figure}[hbt]
\begin{center}
\epsfig{file=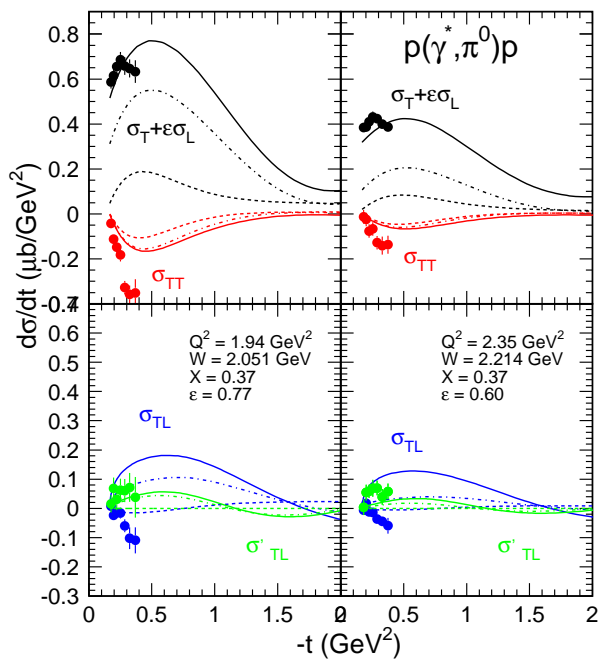,width=3.4in}
\caption[]{(Color on line) The response functions of the reaction $p(\gamma^*,\pi^0)p$ in the two JLab HallA kinematics settings. Dashed lines: pole contributions and Pomeron cut alone. Dash-dotted lines: without $\rho \Delta$ cuts. Full lines: $\rho \Delta$ cuts included.}
\label{HallA_dsdt}
\end{center}
\end{figure}

Fig.~\ref{HallA_dsdt} shows the comparison of the model with the response functions which have been determined at JLab for two values of $Q^2$. I use the following definition:
\begin{eqnarray}
2\pi\frac{d\sigma}{dt d\phi}&=&\frac{d}{dt}\left(\sigma_T +\epsilon \sigma_L
\right. \nonumber \\
&&+\epsilon cos(2\phi) \sigma_{TT} +\sqrt{2\epsilon(\epsilon + 1)} cos(\phi) \sigma_{TL}
\nonumber \\
&&\left. + h \sqrt{2\epsilon(1-\epsilon)} sin(\phi) \sigma'_{TL}
\right)
\label{sigma}
\end{eqnarray}
and I have renormalized the experimental tranverse-longitudinal cross sections~\cite{Cam10}, $\sigma_{TL}$ ad $\sigma'_{TL}$, accordingdly.

\begin{figure}[hbt]
\begin{center}
\epsfig{file=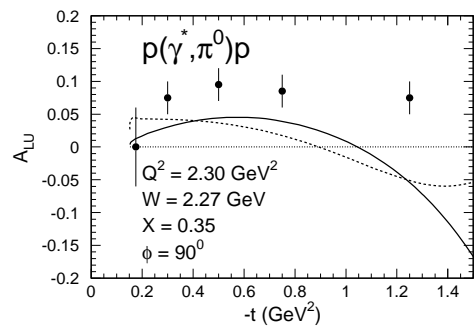,width=2.5in}
\caption[]{ The beam asymmetry~\cite{Ma08} recently measured at JLab, for $Q^2=$  2.3~GeV$^2$, $X=$ 0.35 ($\sqrt{s}=$ 2.27~GeV) and $\phi=$ 90$^0$. The dotted curve takes only into account the $\omega$, $\rho$ and $b_1$ Regge pole contributions. The dash-dotted curve  includes also the CEX cuts contribution. The full curve includes all the cuts contributions.}
\label{ALU}
\end{center}
\end{figure}

Besides the unpolarized cross section ($\sigma_T +\epsilon \sigma_L$), the model reproduces also the Transverse-Transverse interference response function $\sigma_{TT}$. The understanding of the slow variation of the cross section is a non trivial result. At the lowest $Q^2=$ 1.94~GeV$^2$, the mass of the intermediate state is $\sqrt{s}=$ 2.052~GeV, slightly above the $\rho \Delta$ threshold  (2.010~GeV). At fixed $X=$ 0.368,  $\sqrt{s}$ increases when $Q^2$ increases, and the onset of the $\rho \Delta$ cuts compensates the decrease with $Q^2$ and $\sqrt{s}$ of the Regge amplitudes that drive the loop amplitudes.  The contribution of the cuts is small in $\sigma_{TT}$, but large in the Transverse-Longitudinal response functions $\sigma_{TL}$ and $\sigma'_{TL}$. The model predicts the same sign and magnitude as the JLab HallA  fifth response function $\sigma'_{TL}$ (polarized electrons), and therefore reproduces the beam asymmetry $A_{LU}$ (Fig.~\ref{ALU}) that has been recorded in HallB~\cite{Ma08}. But it predicts a different sign for  the Transverse-Longitudinal response function $\sigma_{TL}$. Since it predicts the same sign and magnitude as the preliminary HallB  $\sigma_{TL}$~\cite{Kub10}, I leave open the discussion until the final data are released.

\begin{figure}[hbt]
\begin{center}
\epsfig{file=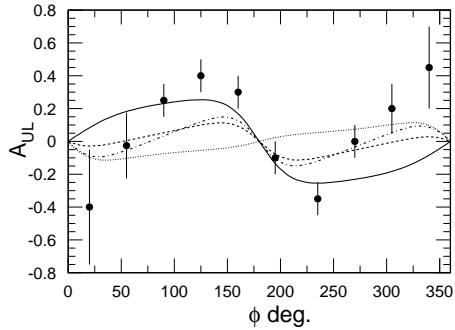,width=2.5in}
\caption[]{ The target asymmetry~\cite{Che06} recently measured at JLab, for $Q^2=$  1.82~GeV$^2$, $\sqrt{s}=$ 2.358~GeV and $t=$ -0.31~GeV$^2$. Dashed line: $\omega$ and $\rho$ poles. Dotted line: $b_1$ pole included. Dash-dotted line: $\pi$ CEX cuts included. Full line: Vector meson cuts included.}
\label{AUL}
\end{center}
\end{figure}

Finally, the unitarity cuts are also the key to the successful interpretation of the target asymmetry $A_{UL}$ (Fig.~\ref{AUL}). The CEX cuts contribute, but falls short. The $\rho^+n$ and $\rho \Delta$ cuts lead to the right sign and the right magnitude of the signal.

In conclusion, the coupling to the vector meson production channels provides us with a natural explanation of the large cross section of the $\pi^0$ electroproduction cross section at $Q^2\sim$ 3~GeV$^2$. Only a few intermediate hadronic states contribute to the unitarity cuts, which are on solid grounds when the elementary production and absorption cross sections are large and are known. So far, we are still in the hadronic regime and we have not reached the domain of factorization between a hard perturbative scattering and a soft non perturbative nucleon structure function.

I acknowledge the warm hospitality at JLab where this work was completed. Jefferson Science Associates operate Thomas Jefferson National Facility for the United States Department of Energy under contract DE-AC05-06OR23177.

\end{document}